\documentclass[12pt,letterpaper]{article}

\usepackage[hmargin={1.5in, 1.5in}, vmargin={1.5in, 1.5in}]{geometry}

\usepackage{rotating}

\usepackage{setspace}

\usepackage{amsmath,amstext,amssymb,amsfonts}
\usepackage{algpseudocode}
\usepackage{algorithm}

\usepackage{subfigure}

\usepackage{paralist}

\usepackage{xcolor,colortbl}

\renewcommand{\Pr}{\mathsf{P}}

\newtheorem{theorem}{Theorem}[section]
\newtheorem{proposition}[theorem]{Proposition}

\usepackage{graphicx}

\usepackage{natbib}

\begin{document}


\title{BayesCAT: Bayesian Co-estimation of Alignment and Tree}
\date{}









\author
{Heejung Shim\\
shim@gmail.com \\
Department of Human Genetics, University of Chicago\\
Bret Larget\\
brlarget@wisc.edu\\
Departments of Statistics and of Botany, University of Wisconsin, Madison}

\maketitle









\begin{abstract}
Traditionally, phylogeny and sequence alignment are estimated separately: first estimate a multiple sequence alignment and then infer a phylogeny based on the sequence alignment estimated in the previous step. However, uncertainty in the alignment estimation is ignored, resulting, possibly, in overstated certainty in phylogeny estimates. We develop a joint model for co-estimating phylogeny and sequence alignment which improves estimates from the traditional approach by accounting for uncertainty in the alignment in phylogenetic inferences. Our insertion and deletion (indel) model allows arbitrary-length overlapping indel events and a general distribution for indel fragment size. We employ a Bayesian approach using MCMC to estimate the joint posterior distribution of a phylogenetic tree and a multiple sequence alignment. Our approach has a tree and a complete history of indel events mapped onto the tree as the state space of the Markov Chain while alternative previous approaches have a tree and an alignment. A large state space containing a complete history of indel events makes our MCMC approach more challenging, but it enables us to infer more information about the indel process. The performances of this joint method and traditional sequential methods are compared using simulated data as well as real data. Software named BayesCAT (Bayesian Co-estimation of Alignment and Tree) is available at https://github.com/heejungshim/BayesCAT. 
\end{abstract}

\section{Introduction}
\label{s:intro}
The use of molecular sequences is popular in phylogeny estimation and a large number of statistical methods have been proposed. 
Phylogenetic inference using molecular sequences traditionally consists of two separate steps: first estimate a multiple sequence alignment and then infer a phylogeny based on the sequence alignment estimated in the previous step.
This sequential approach ignores alignment uncertainty, leading to several problems with phylogenetic inference. 

If the alignment contains ambiguous regions, ignoring uncertainty in the alignment can result in exaggerated support for inferred phylogenies \citep{Lutzoni2000}. 
Moreover, if the sequence alignment is determined by an alignment method that assumes a fixed guide tree, then the estimated phylogeny in the second step may be biased toward this fixed guide tree \citep{Lake1991, Thorne92, Sinsheimer94, Nelesen2008}. 
As various alignment methods typically align ambiguous regions differently, phylogenies estimated by the traditional sequential approach can change considerably according to the choice of aligment program.
\citet{Wong08} show that different alignment methods can lead to different conclusions in a comparative genomics study.
(See Web Appendix A for our investigation of the problems of the traditional sequential approach using a simulated data set).
A simple approach to avoid these problems is to exclude ambiguous regions in the following phylogeny estimation procedure.
However, the decision of which regions are ambiguous is subjective and ambiguous regions can include a large fraction of potentially informative sites \citep{Lutzoni2000}.
Another approach to sidestep the limitations of the traditional sequential approach is to estimate alignment and phylogeny simultaneously. Thus, researchers have developed diverse methods for joint estimation of alignment and phylogeny including statistical approaches \citep{Lunter05, Redelings05, Redelings07, Novak2008} and non-statistical approaches \citep{VarAn2010, SATE, Liu2012}. More comprehensive background on these methods can be found in \citet{Redelings05}. 

Statistical approaches to joint estimation use mutation rates, insertion and deletion (indel) rates, and divergence time, instead of penalties (e.g., gap and mismatch penalties) employed in most non-statistical methods, so that these approaches take into account multiple occurrences of mutations and indels at each site. 
In addition, statistical approaches use statistical models for the substitution process and the indel process, allowing for inferences about the nature of the process of evolution. 
In particular, Bayesian statistical approaches provide a framework to measure uncertainty in the estimated alignment and tree. 

\citet{Lunter05} developed a fully Bayesian method which uses the TKF91 model \citep{TKF91} for indel events. 
The TKF91 model has the restriction of allowing only single-base indels. 
This restriction tends to overemphasize the information in a single long indel by treating one event as many, which can affect posterior estimates \citep{Redelings05}. 
Redelings and Suchard also proposed a Bayesian approach (BAli-Phy). 
In their first paper \citep{Redelings05}, they allow indels to contain a geometrically distributed number of bases. 
Although their model does not allow indels on the same branch to overlap, it improves on the TKF92 model \citep{TKF92} by avoiding a fragment-based indel process. 
However, this improvement was made possible by assuming that occurrence of indel events on each branch is independent of branch length.
As it is biologically reasonable to expect more indel events on longer branches, this assumption is undesirable.
Their second paper \citep{Redelings07} removed the assumption, which results in a fragment-based indel model like the TKF92 model on individual branches. 
\citet{Novak2008} also developed a software package where they used the long indel model introduced in \citet{longIndel03}. 
The long indel model improves on the TKF91 and TKF92 models by allowing indels to have multiple bases and overlap. 
\citet{longIndel03} introduce an algorithm for calculation of alignment likelihoods under the long indel model, but it is based on approximation by bounding the number of indel events and the indel fragment size per event.

The statistical joint estimation methods above sum over all possible indel histories under their models, which yields a restricted inference on the indel process itself.
Estimated multiple alignments show inferred homologies, but are not easy to interpret with regard to specific indel event histories. 
In addition, to achieve this summability, the models disallow many biologically plausible indel histories.

In this paper, we develop a model for joint estimation of alignment and phylogeny and design MCMC methods to carry out Bayesian inference.
We propose an indel model which allows arbitrary-length overlapping indel events and a general distribution for indel fragment size. 
We use the exact likelihood of the indel history under our indel model instead of an approximated likelihood of alignment.
The major difference between our approach and the previous approaches to the joint estimation of tree and alignment is the state space of the Markov Chain. 
Our approach has a tree and a complete history of indel events on the tree as the state space while the previous approaches have a tree and an alignment. 
A large state space containing a complete history of indel events makes our MCMC approach more challenging, but it enables us to infer more information about the indel process. 

\section{Model}
\label{ch:model}
To model the evolution of molecular sequences, we consider the process of nucleotide substitution in which single sites change bases and the indel process in which DNA fragments are inserted into or deleted from the sequence. In our joint model for co-estimation of phylogeny and alignment, these two processes can be separated, and moreover, the traditional substitution models used with fixed alignments can be adopted. In this paper, we develop an indel model, which allows arbitrary-length overlapping indel events and a general distribution for indel fragment size.
\subsection{Joint model}
\label{model:joint}
The observed data $S$ consists of $n$ unaligned sequences. The $n$ unaligned sequences are related by a phylogenetic tree $T$ and aligned by a history of indel events $H$ on the tree. 
The phylogenetic tree $T$ is composed of an unrooted bifurcating tree topology $\tau$ and branch lengths denoted as $V = (v_1, \ldots, v_{2n-3})$. The indel history $H$ includes for each edge a sequence of events which consist of the time, type (insertion or deletion), position on the sequence, and inserted or deleted fragment size (see Section~\ref{model:introIndelP} and Section~\ref{model:indelNotation} for details). The unnormalized posterior distribution of the tree $T$ and indel history $H$ given the observed sequences $S$ is:
\begin{eqnarray*}
\Pr(H, \tau, V \mid S, \Theta) &\propto& \Pr(S \mid H, \tau, V, \Theta_{\text{sub}})\Pr(H \mid \tau, V, \Theta_{\text{ID}})\Pr(\tau, V \mid \Theta_{\text{tree}})
\end{eqnarray*}
where $\Theta$ consists of three components, $\Theta_{\text{sub}}$, $\Theta_{\text{ID}}$, and $\Theta_{\text{tree}}$, for the nucleotide substitution process, indel process, and the tree, respectively. On the right-hand side of the equation, the first factor is the likelihood of the sequences and is given by a substitution model. The second factor, the probability of the indel history on a given tree is specified by our indel model, which will be described in detail later. For the third factor, we assume a uniform distribution over unrooted tree topologies with $n$ taxa and independent exponential distributions with common mean $1/\gamma$ on the length of each branch, leading to this expression:
\begin{eqnarray*}
\Pr(H, \tau, V \mid S, \Theta) &\propto& \Pr(S \mid H, \tau, V, \Theta_{\text{sub}})\Pr(H \mid \tau, V, \Theta_{\text{ID}})\Pr(\tau)\Pr(V \mid \Theta_{\text{tree}})
\end{eqnarray*} 
where $\Theta_{\text{tree}} = \gamma$. 
\subsection{Substitution model}
\label{model:sub}
A history of indel events $H$ on a tree $T$ determines a multiple alignment $A$ (up to minor reordering of some columns) where homologous residues are aligned in columns. 
Different indel histories might give rise to the same alignment. The alignment $A$ is sufficient for the substitution process, yielding $\Pr(S \mid H, \tau, V, \Theta_{\text{sub}}) = \Pr(S \mid A, \tau, V, \Theta_{\text{sub}})$, where $A$ is a function of $H$ on $T = (\tau, V)$. 
The traditional substitution models used with fixed alignments can be adopted here since the right-hand side of the equation has the same form. We assume substitutions occur independently across columns of $A$ according to a continuous-time Markov process. 
At present, we only consider reversible Markov models, which leads to the use of an unrooted tree topology. Any substitution model could be used. 

We use the HKY model \citep{Hasegawa1985} in our analysis here (see Web Appendix B for details of the HKY model), so $\Theta_{\text{sub}}$ consists of $\kappa$, the ratio of the transition to transversion rates among nucleotides, and nucleotide frequencies in the equilibrium distribution of the rate matrix, denoted as $\pi = (\pi_A, \pi_C, \pi_G, \pi_T)$. 

\subsection{Indel model}
\label{model:indel}

\subsubsection{Description of an indel history on a tree}
\label{model:introIndelP}
To help our description of indel models, this section first describes indel events for a given sequence, and then illustrates indel events on a tree using simple examples. A molecular sequence is represented by a sequence of symbols with one symbol for each base in the sequence. 
We number the bases from left to right in a sequence of length $n$ from one to $n$. 
We use the term {\it position} to refer to the places between or at the ends of bases in a sequence where indel events might act.
A sequence of length $n$ has $n+1$ positions which we number from zero to $n$ from left to right, so base $i$ is between positions $i-1$ and $i$.
Deletions remove all bases between two positions. 
We identify a deletion event with the leftmost position and the number of bases deleted.
In general, a deletion event of size $x$ at position $i$ removes all bases between positions $i$ and $i+x$.
Insertions act at a single position by adding one or more bases to the sequence. See Web Figure 1 for examples of indel events for a given sequence.
\begin{figure}
\centering
\includegraphics[scale=0.8]{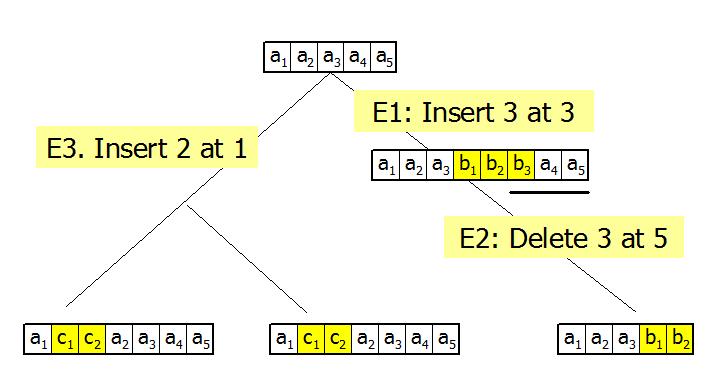}
\caption[{\bf Example : A complete history of indel events on a tree.}]{\label{indelHonTree:fig} {\bf Example : A complete history of indel events on a tree. }
{\small Three indel events occur on a rooted tree, in which a sequence at the root has five residues. Event $E1$ on the right child edge inserts three residues, colored yellow, at position three into the root sequence. Event $E2$, which occurs after $E1$ on the same edge, deletes three residues at position five. The deleted residues are underlined. The rightmost leaf retains five residues. Event $E3$ on the left child edge is independent of $E1$ and $E2$. It inserts two residues at position one in the root sequence. The remaining two leaves have sequences of length seven. Residues with the same notation are homologous.}}
\end{figure}

We specify a complete history of indel events on a tree relative to rooting the tree at one time point.
As depicted in Figure~\ref{indelHonTree:fig}, we represent an insertion or deletion event on the rooted tree based on the sequence before the occurrence of that event.
Events $E1$ and $E3$ are specified based on the root sequence and event $E2$ is represented relative to the sequence after event $E1$. 
A history of indel events on a tree determines {\it homologous} residues, i.e., residues derived from a common ancestor. 
In Figure~\ref{indelHonTree:fig}, residues with the same notation are homologous.

\subsubsection{General indel model}
\label{model:indelModel}  
We develop a general indel model that allows arbitrary-length overlapping indels and a general distribution for the indel fragment size.
We imagine a sequence of interest embedded within a much longer sequence which undergoes a homogeneous process of insertion and deletion, conditional on leaving the endpoints of the sequence of interest intact. 

Our indel model rests on the following assumptions: 
(1)~time reversibility; 
(2)~insertions can occur at any positions on a given sequence including the end positions; 
(3)~insertion fragments can be of any size; 
(4)~for a fragment of a given size, the insertion rate is spatially homogeneous on the sequence;
(5)~deletions can occur at any positions on the sequence except for the end position; 
(6)~for a given position on the sequence, the deletion fragment has maximum size, which is the number of residues to the right of the position;
(7)~for a fragment of a given size, the deletion rate is spatially homogeneous over possible positions on the sequence; 
(8)~non-zero deletion rate of a single-residue; and (9)~the total insertion rate and total deletion rate per site on the sequence are finite. 

We specify the constant insertion and deletion rates for a fragment of $k$ bases using $\lambda i(k)$ and $\mu d(k)$, respectively, where $d(j) \ge 0$ and $i(j) \ge 0$ for all $j \in \{ 1, 2, \ldots\}$, $\sum_{j=1}^{\infty}i(j) = 1$, and $\sum_{j=1}^{\infty}d(j) = 1$.
Then, $\lambda$ and $\mu$ represent the total insertion and deletion rates per site on a sequence of infinite length, respectively. 
In addition, $i(\cdot)$ and $d(\cdot)$ have the same meaning as the insertion and deletion fragment size distributions on the sequence of infinite length, respectively.  
In this paper, we call $i(\cdot)$ and $d(\cdot)$ the base insertion and deletion fragment size distributions, respectively. 

To clarify these assumptions, we present several comments.
First, assumption~(6) implies deletions have no possible fragment size at the right end of the sequence, which is the reason we exclude this position from possible positions for deletion in assumption~(5).
Second, for a given fragment size, deletions have a restriction on possible positions of the sequence due to assumption~(6). 
Third, under the assumptions above, the total insertion rate per site is homogeneous, but the total deletion rate per site depends on the position on the sequence.
Due to assumption~(6), the total deletion rate per site decreases as the site approaches the right end of the sequence (see the example in Web Figure 2).
Fourth, the insertion fragment size distribution is identical to $i(\cdot)$ at all positions.  
However, due to assumption~(6), the deletion fragment size distribution at a given position depends on the position on the sequence and it is truncated distribution of $d(\cdot)$ (see the example in Web Figure 2).
Finally, assumption~(8) is required in the derivation of our model, but allowing indel events of unit base is also biologically realistic.

The components of a general indel model that follows the previous assumptions include the equilibrium length distribution $q(\cdot)$, base indel fragment size distributions $i(\cdot)$ and $d(\cdot)$, and indel rates $\lambda$ and $\mu$. 
The following proposition describes the most general indel model that satisfies the above assumptions.
\begin{proposition}
\label{indelThm}
Under the previous assumptions, the equilibrium length distribution $q(\cdot)$ is
\begin{eqnarray*}
q(x) &=& r(1-r)^x \mbox{ for all $x$} \in \{ 0, 1, \ldots \},
\end{eqnarray*}  
where $1-r = \frac{{\lambda}i(1)}{{\mu}d(1)}$ and $0 < r < 1$.
The base deletion fragment size distribution $d(\cdot)$ can be any distribution with support on the positive integers and $d(1) > 0$, and the ratio of the insertion rate to the deletion rate is
\begin{eqnarray*}
\frac{\lambda}{\mu} &=& \sum_{k=1}^{\infty}{(1-r)^kd(k)} < 1.
\end{eqnarray*}  
The base insertion fragment size distribution $i(\cdot)$ is determined as
\begin{eqnarray*}
i(k) &=& \frac{\mu}{\lambda}(1-r)^kd(k)\mbox{ for all $k$} \in \{ 1, 2, \ldots\}.
\end{eqnarray*} 
\end{proposition}
The proof is given in Web Appendix C. The most general indel model allowed under the assumptions above permits free specification of the parameter $r$, the distribution $d(\cdot)$, and one of $\lambda$ or $\mu$, but the remaining components of the model are then determined. 

\subsubsection{Examples of general indel model}
\label{model:indelExamples}
The selection of a particular distribution for the deletion fragment size determines examples of the general indel model. We illustrate a geometric distribution here and negative binomial and power law distributions in Web Appendix D. We consider a geometric distribution with parameter $r_d$ as the deletion fragment size distribution $d(\cdot)$. 
Then, the insertion fragment size distribution $i(\cdot)$ is also geometric as follows.
Selection of $0 < r, r_d < 1$ yields $q(x) = r(1-r)^x$ for $x = 0, 1, \ldots$, $d(k) = r_d(1-r_d)^{(k-1)}$ and $i(k) = r_i(1-r_i)^{(k-1)}$ for $k = 1, 2, \ldots$, and $\frac{\lambda}{\mu} = \frac{r_d(1-r_i)}{r_i(1-r_d)}$, where $r_i = 1 - (1-r_d)(1-r)$, $0 < r, r_d, r_i < 1$, $r_i > r_d$, $r_i > r$, and $\mu > \lambda > 0$. 
As the choice of $\lambda > 0$ determines $\mu$ or vice versa, this model has three free parameters. 
That is, $\Theta_{\text{ID}} = (r, r_d, \lambda)$. 
As insertion has one more possible position than deletion on a sequence, the constraint $\mu > \lambda$ is necessary in our model to prevent sequences from growing indefinitely over time. 
The requirement $r_i > r_d$ is also reasonable because for a given position on the sequence, the deletion fragment size has a maximum possible value while the insertion fragment size is not restricted. 

\subsubsection{Relationship with the previous indel models}
\label{model:indelOthers}
Allowing $r_d = 1$ under the geometric model results in the TKF91 model as a special case of our model. 

It turns out that our indel model is very similar to the long indel model \citep{longIndel03}. 
Under our indel model, for a fragment of a given size, the insertion rate and the deletion rate per site are spatially homogeneous over possible positions on the sequence.
The total insertion rate per site is also homogeneous. 
However, the total deletion rate per site depends on the position on the sequence because the deletion fragment has maximum size, which is the number of residues on the right hand side of the position. 
Thus, the total deletion rate per site decreases as the site approaches the right end of the sequence in our indel model. 
Conversely, the long indel model assumes that the total insertion rate and the total deletion rate per site are spatially homogeneous, which leads to increased insertion and deletion rates of a given fragment size at both ends of the sequence.
\citet{longIndel03} introduce an algorithm for calculation of alignment likelihoods under the long indel model, but it is based on approximation by bounding the number of indel events and the indel fragment size per event.
In this paper, we use the exact likelihood of the indel history under our indel model instead of an approximated likelihood of alignment.

\subsubsection{Specific description of an indel history on a single edge}
\label{model:indelNotation}
Although the tree in our model is unrooted, we assume a time direction for convenience when calculating the likelihood or updating an indel history on the tree. 
Thus, we will describe specific components of an indel history on a single edge and introduce their notation after assuming the single edge has defined parent and child nodes.  
Let an indel history $h$ on a single edge of length $v$ have $K$ indel events. 
These events are ordered by their occurrence time on the edge defined relative to the parent node. 
In the $i$th event $e_i = (t_i, id_i, p_i, l_i, n_i)$, $t_i$ indicates the time of this event defined as a distance from the parent node to the event; $id_i$ denotes its type, whether the event is an insertion or deletion; $p_i$ signifies the position on the sequence where a fragment for deletion starts or a new fragment is inserted; $l_i$ is the size of the inserted or deleted fragment; $n_i$ is the total length of the sequence after the $i$th event. For convenience, let $n_0$ be the sequence length at the parent node and let $n_{K+1} = n_K$ be the sequence length at the child node.
Then, $n_i = n_{i-1} + l_i$ if the $i$th event is an insertion and $n_i = n_{i-1} - l_i$ if the $i$th event is a deletion.
Let $t_0 = 0$ and $t_{K+1} = v$, which is the length of the single edge.

\subsubsection{Indel history probability density calculation} 
\label{model:indelLike}

We first derive the likelihood for the history on a single edge and then for the entire tree.

\paragraph{On a single edge} Under our indel model, the likelihood for an indel history $h = (e_1, e_2, \ldots, e_K)$ on a single edge, conditional on the branch length $v$ and the sequence length of the parent node $n_0$, is computed as the product of exponentially-distributed waiting times for each event multiplied by an exponential tail probability for no further events in the remaining interval.
\begin{eqnarray*}
\Pr(h \mid v, n_0) = \left[\prod_{j=1}^K \Pr(e_j \mid t_{j-1}, n_{j-1})\right]\exp(-\eta_{K+1}(t_{K+1}-t_{K})),
\end{eqnarray*}  
where $\eta_j = (n_{j-1} + 1)\lambda + f(n_{j-1})\mu$ is the total intensity of indel rates across all positions and $f(x) = \sum_{k=1}^{x}(x-k+1){d(k)}$
is used to sum deletion probabilities over all positions and all allowable deletion sizes.
The probability density of each event involves choosing the time given the current time and length,
the type of event (insertion or deletion) given the current length and that the event occurs,
and the position and size of the event given its type and the current length.
\begin{eqnarray*}
\Pr(e_j \mid t_{j-1}, n_{j-1}) = \Pr(p_j, l_j \mid id_j, n_{j-1})\Pr(id_j \mid n_{j-1})\Pr(t_j \mid t_{j-1}, n_{j-1}),
\end{eqnarray*}  
where $\Pr(t_j \mid t_{j-1}, n_{j-1}) = \eta_{j}\exp(-\eta_j(t_j-t_{j-1}))$ and 
\begin{displaymath}
\Pr(id_j \mid n_{j-1}) = \left\{ \begin{array}{ll}
   \frac{(n_{j-1} + 1)\lambda}{\eta_j} & \text{if   } id_j = \mathrm{in}\\
   \frac{f(n_{j-1})\mu}{\eta_j} & \text{if   } id_j = \mathrm{del}.
\end{array} \right. 
\end{displaymath}
If $id_j = \mathrm{in}$, for $p_j \in \{0,1, \ldots, n_{j-1}\}$ and $l_j \in \{1, 2, \ldots\}$,
then $\Pr(p_j, l_j \mid id_j, n_{j-1}) = \frac{i(l_j)}{n_{j-1} + 1}$.
If $id_j = \mathrm{del}$, for $p_j \in \{0,1, \ldots, n_{j-1}-1\}$ and $l_j \in \{1, \ldots, n_{j-1}-p_j\}$,
then $\Pr(p_j, l_j \mid id_j, n_{j-1}) = \frac{d(l_j)}{f(n_{j-1})}$.
Putting this together,
\begin{displaymath}
\Pr(e_j \mid t_{j-1}, n_{j-1}) = \left\{ \begin{array}{ll}
   \exp(-\eta_j(t_j-t_{j-1}))\lambda i(l_j) & \text{if   } id_j = \mathrm{in}\\
   \exp(-\eta_j(t_j-t_{j-1}))\mu d(l_j) & \text{if   } id_j = \mathrm{del}.
\end{array} \right. 
\end{displaymath}
Therefore, the probability density for an indel history on a single edge simplifies to 
\begin{equation}\label{singleL}
\Pr(h \mid v, n_0) = \exp \left(-\sum_{j=1}^{K+1}\eta_j(t_j-t_{j-1})\right)\prod_{j=1}^K\left[(\lambda i(l_j))^{I_{\{id_j = \mathrm{in}\}}} (\mu d(l_j))^{I_{\{id_j = \mathrm{del}\}}}\right].
\end{equation}   

\paragraph{On the tree} For a given tree $T = (\tau, V)$ with $n$ taxa, let $V = (v_1, \ldots, v_{2n-3})$ and $H = (h_1, \ldots, h_{2n-3})$ refer to the branch lengths and the indel histories of the edges. 
For convenience in calculation of the probability of an indel history on a tree, we assume one node of the tree $T$ is a root, and then define an artificial parent node for each edge be the node nearest to this root.
Let $h_j$ represent the indel history on the $j$th edge under the defined direction and $n_j$ denote the sequence length of its parent node.
Then, the probability density for the indel history $H$ on the tree $T$ is $\Pr(H \mid T) = q(n_r)\prod_{j=1}^{2n-3}\Pr(h_j \mid v_j, n_j)$, where $n_r$ indicates the sequence length at the root and $q(\cdot)$ represents the equilibrium length distribution. 
The probability density $\Pr(h_j \mid v_j, n_j)$ is calculated by formula (\ref{singleL}) for the $j$th edge.

\subsection{Specification of the prior distributions}
\label{model:Prior}
The parameters are partitioned into three parts, each with an independent prior distribution.

\paragraph{Branch lengths}
$\Theta_{\text{tree}} = \gamma$: For a prior on $\gamma$, we assume the density $g(\gamma) = \frac{\alpha_{\gamma}}{{(1 + \alpha_{\gamma}\gamma)}^2}$ for $\gamma > 0$, which arises as a ratio of independent exponential random variables and has a very heavy tail; the mean is infinite, but the median is $1/\alpha_{\gamma}$.

\paragraph{Substitution model}
$\Theta_{\text{sub}} = (\pi, \kappa)$: We assume a Dirichlet prior distribution with parameters $\alpha_{\pi} = (\alpha_A,\alpha_C,\alpha_G,\alpha_T)$ for $\pi$ and a ratio of exponentials prior distribution with a parameter $\alpha_{\kappa}$ for $\kappa$.

\paragraph{Indel model}
$\Theta_{\text{ID}} = (r, r_d, \lambda)$: We assume a beta prior distribution with parameters ($\alpha_r$, $\beta_r$) and ($\alpha_{r_d}$, $\beta_{r_d}$) for $r$ and $r_d$, respectively, and an exponential prior distribution with a parameter $\alpha_{\lambda}$ for $\lambda$. 

\section{MCMC approach}
\label{se:MCMC}
We sample from the joint posterior distribution $P(H, \tau, V, \Theta \mid S)$ using MCMC to estimate the alignment, tree, and model parameters and to quantify uncertainty in these estimates. 
To sample from the entire state space containing a tree $T$, an indel history $H$ on the tree, and model parameters $\Theta$, we use several MCMC updates employing a random-scan line \citep{Liu1995}, Metropolis-within-Gibbs \citep{Tierney} approach. 
MCMC using reversible jump \citep{Green1995} is adopted in updates involving the indel history due to changes in the dimension of the state space. 
Our MCMC proposal methods have four categories (overview of the proposal methods in these four categories is shown in Web Appendix E).
The proposal in the first category updates the branch length ($V$) of a randomly selected edge. 
Although the times of the indel events ($H$) on the edge change in proportion to the change of the edge length, this update method does not vary alignments. 
Proposal methods in the second category select an edge of the tree at random and propose a new indel history ($H$) on the edge, conditional on the fixed sequence lengths of the two nodes connected by the edge. 
Here, the proposed new history can modify the alignment of sequences ($A$). 
Proposal methods in the third category pick an internal node and update an indel history ($H$) on three edges, which are adjacent to the internal node. This method updates an alignment ($A$), a sequence length at the internal node, and branch lengths ($V$) of the edges adjacent to the internal node. 
The last category contains proposal methods of subtree pruning and regrafting which update a tree topology ($\tau$), a sequence length at an internal node, an indel history ($H$), an alignment ($A$), and the collection of branch lengths ($V$). 

As part of the effort made to validate the implementation of our MCMC methods, we generate many data sets from the prior distribution, run MCMC on each one, calculate summary statistics of interest from each sample, and average these across samples. Close agreement between these results and expected values from the prior distribution is evidence of correct derivation and implementation of our MCMC apporach (see Web Appendix F for detailed procedure and results). 

\citet{Shimthesis} describes all the update methods in detail. 
The proposals for changing the tree and substitution model parameters are common in the Bayesian phylogenetics literature.
The proposals that modify an indel history on an edge are novel to the modeling approach in this paper
and are incorporated into the proposals that modify other parts of the parameter space as well.
Here, we describe in detail algorithms for proposing an indel history on a single edge, conditional on the sequence lengths of the two nodes connected by the edge being fixed. 

\subsection{Propose a new indel history on a single edge}
\label{method:EdgeBasic}
For a given edge of length $v$ with parent and child nodes of sequence lengths $n_0$ and $n_v$, respectively, we propose new indel history $h$.
A provisional history is generated sequentially starting from the sequence at the parent node using the Markov model for the indel process. 
The time to the next event is generated from an exponential distribution whose rate is the total sum of the rates of all possible next events on the current sequence.
An insertion or deletion event is proposed according to its rate on the current sequence. 
The sequence length changes after each indel event. 
This process proceeds until the next event time exceeds the length of the edge.
If the length of the final sequence differs from $n_v$, one additional event is appended to the provisional history at a random time between the last event and $v$ with the type and fragment size chosen to match the required sequence length at the end.
The detailed proposal algorithm is provided in Web Appendix G. 

The probability of proposing a indel history $h = (e_1, e_2, \ldots, e_K)$ under this procedure, $Q(h \mid v, n_0, n_v)$, is calculated as follows. If $K > 0$, 
\begin{eqnarray*}
Q(h \mid v, n_0, n_v) = \left[\prod_{i=1}^{K-1} \Pr(e_i \mid t_{i-1}, n_{i-1})\right]\Pr(e_K \mid t_{K-1}, n_{K-1}, v, n_v).
\end{eqnarray*}
Define $\eta_i$ and $f(x)$ as above and let $q_{\text{in}}(\cdot)$ and $q_{\text{del}}(\cdot)$ be the probabilities of proposing an insertion and a deletion of a given size, respectively. Then, 
\begin{displaymath}
\Pr(e_i \mid t_{i-1}, n_{i-1}) = \left\{ \begin{array}{ll}
   \exp(-\eta_i(t_i-t_{i-1}))\lambda q_{\text{in}}(l_i) & \text{if   } id_i = \mathrm{in}\\
   \exp(-\eta_i(t_i-t_{i-1}))\frac{f(n_{i-1})\mu q_{\text{del}}(l_i)}{n_{i-1}-l_i+1} & \text{if   } id_i = \mathrm{del}
\end{array} \right. 
\end{displaymath}
and $\Pr(e_K \mid t_{K-1}, n_{K-1}, v, n_v)$ is 
\begin{displaymath}
\left\{ \begin{array}{ll}
\exp(-\eta_K(t_K-t_{K-1})-\eta_{K+1}(v-t_K))\lambda q_{\text{in}}(l_K)+\frac{\exp(-\eta_K(v-t_{K-1}))}{(v-t_{K-1})(n_{K-1}+1)} & \text{if   } id_i = \mathrm{in}\\
\exp(-\eta_K(t_K-t_{K-1})-\eta_{K+1}(v-t_K))\frac{f(n_{K-1})\mu q_{\text{del}}(l_K)}{n_{K-1}-l_K+1}+\frac{\exp(-\eta_K(v-t_{K-1}))}{(v-t_{K-1})(n_{K-1}-l_K+1)} & \text{if   } id_i = \mathrm{del}.
\end{array} \right. 
\end{displaymath}
If there are no events ($K=0$), then $Q(h\mid v, n_0, n_v) = \exp(-\eta_1v)$.

\subsubsection{Propose a new indel history on a single edge considering the sequence length at the child node} 
\label{method:EdgeChild}
The proposal introduced above takes into account the sequence length at the child node ($n_v$) only at the last step when proposing one additional event. 
This can lead to a high probability of proposing unlikely histories that are longer than more likely histories.
For instance, if a sequence at a parent node has two more bases than a sequence at a child node, a single deletion event with a fragment size of two bases might be the most probable history. 
However, about half of the proposed histories will begin with an insertion event, and a further deletion event will be required. 
An alternative proposal method includes these modifications: (1)~an increased probability of proposing no additional events when the current sequence length matches the target; (2)~an increased probability of proposing an insertion (deletion) when the target length is greater (less) than the current length; and (3)~an increased probability of proposing a fragment size to match the target sequence length.  
Although this proposal introduces a number of tuning parameters and comparison steps, we observe that it helps to increase MCMC mixing. 
The detailed description and proposal probability are provided in Web Appendix H.

\subsection{Alignment summary}
\label{method:AlignSummary}
To summarize samples of alignments, we present an alignment with maximal expected accuracy and visualize uncertainty for every column and character of the alignment with color, which is accomplished using the method proposed by \citet{FSA} and implemented in the program FSA (Fast Statistical Alignment). 

The software FSA consists of two separate parts. The first part of FSA performs pairwise comparisons of the input sequences to estimate the posterior probabilities that individual characters are aligned using the standard three- or five-state pair hidden Markov model \citep{HMM}. The second part of FSA constructs a multiple alignment from the posterior probabilities estimated at the first part using a sequence annealing technique \citep{annealing}. This procedure produces the multiple alignment with maximal expected accuracy, which is defined as a multiple alignment with minimal expected distance to the true alignment. The true alignment is treated as a random variable whose distribution is determined under a statistical model used in the first step.

Instead of the first step of FSA, we estimate the posterior probabilities for each pair of sequences from our multiple alignment samples. Then, we adopt the second part of FSA to construct the multiple alignment with maximal expected accuracy. Since the posterior probabilities used in the second step of FSA are estimated under our model, the final multiple alignment has maximal expected accuracy under our model (see Web Appendix I).
Figure~\ref{5srRNAalign:fig} shows an example of alignment summarization. Each character (gap) is colored according to the expected accuracy with which each character (gap) is aligned to other characters or gaps in the column. We note that FSA allows alignment uncertainty to be evaluated by other measurements : sensitivity, specificity, certainty, and consistency.

\section{Applications}
\label{se:Results}
We apply our approach to joint estimation of the alignment and tree (BayesCAT) to a data set from \citet{Redelings05}, and then compare the performance of BayesCAT with the traditional sequential methods and with an alternative joint model approach, BAli-Phy \citep{BAli}. In addition, we conducted a comparison on simulated data, where the true tree, indel history, and alignment are known, and provide a detailed procedure and comparison results in Web Appendix J.

\subsection{Data description: 5S rRNA}
\label{Results:5srRNAdata}
A question of interest in \citet{Redelings05} is whether the Archaea form a monophyletic group, one of the important unresolved question about deep branches in the Tree of Life \citep{Brown1997}. 
\begin{table}[t]
\caption[{\bf 5S rRNA : Data description.}]{\label{5srRNAdata:table}\textbf{5S rRNA : Data description. } Note that boldface here and in other tables represents Archaea species. Abbreviation corresponding to each species is shown in parentheses.}
  \begin{center}
    \begin{tabular}{|c|c|c|}
      \hline
      Taxa&Domain&Order\\ \hline \hline
      Escherichia coli (EC) & Bacteria&Proteobacteria\\\hline
      Homo sapiens (HS) & Eukaryotes& Metazoa\\\hline
      Halobacterium salinarum {\bf (HA)} & {\bf Archaea}& Euryarchaeota\\\hline
      Pyrococcus woesei {\bf (PW)} & {\bf Archaea}&Euryarchaeota\\\hline
      Sulfolobus acidocaldarius {\bf (SA)} & {\bf Archaea}&Crenarchaeota \\\hline
    \end{tabular}
  \end{center}
\end{table}   

We begin with a brief summary of the background introduced in \citet{Redelings05}. 
The division of all living organisms into the three domains (Archaea, Bacteria, and Eucarya) was proposed by \citet{Woese1990} and has been supported by research into the molecular biology of Archaea \citep{Brown1997}. 
\citet{Woese1990} suggest that Archaea form a monophyletic group, but some other analyses suggest that the Crenarchaeotes, also called Eocytes after \citet{Lake1991}, separated from the remaining Archaea and form a clade with the Eukaryotes \citep{Rivera1992}.
These conflicting results suggest two alternative hypotheses about the early branching in the Tree of Life. 
The archaea tree represents the hypothesis that the Archaea form a monophyletic group while the eocyte tree denotes the alternative hypothesis that the eocyte Archaea are more closely related to Eukaryotes than to the remaining Archaea (See Figure 5 in  \citet{Redelings05}).

The 5S rRNA, a component of the large ribosomal subunit, is found in Archaea, Bacteria, and Eukaryotes and has a highly conserved secondary structure \citep{Barciszewska}. The 5S rRNA sequences in the data set range in length from 120 to 126 base pairs and the lowest pairwise sequence identity is around 46$\%$ \citep{Redelings05}. Table~\ref{5srRNAdata:table} lists five taxa used in the analysis. \citet{Redelings05} also used this data to compare BAli-Phy to the traditional sequential approach. 

\subsection{Model and prior distributions}
\label{Results:5srRNAmodel}
We use the HKY model \citep{Hasegawa1985} in our analysis here (see Web Appendix B for details of the HKY model), and we use the geometric distribution for the deletion fragment size. The prior distribution of the remaining parameters are described in Section~\ref{model:Prior}.
We assume a Dirichlet prior distribution with parameters $\alpha_{\pi} = (13.3, 21.7, 23.1, 11.9)$ for $\pi$. The parameter $\alpha_{\pi}$ is selected to have the observed frequencies of bases as a mean and to cover broad regions. 
We assume a beta prior distribution with parameters (100, 12200) and (3, 15) for $r$ and $r_d$, respectively. 
These prior distributions are selected to cover reasonably broad regions based on the observed sequence lengths. 
The posterior estimates of each parameter, together with a prior mean (and median for $\gamma$ and $\kappa$), are summarized in Web Table 1.  

\subsection{Phylogeny estimation}
\label{Results:5srRNAtree}
Table~\ref{5srRNAtree:table} shows posterior probabilities of the top three topologies, ranked by posterior probabilities from BayesCAT. 
No single topology is strongly supported by BayesCAT as the support for the most probable topology is only 0.205.  
The posterior probabilities for the archaea tree ($T2$) and the eocyte tree (not shown) are 0.17 and 0.078, respectively.
The split, which supports the hypothesis of archaeal monophyly, has a posterior probability of 0.414 (Table~\ref{5srRNAtree:table}). 

Table~\ref{5srRNAtree:table} also lists posterior probabilities from the traditional sequential approach where we apply MrBayes to the alignment determined using ClustalW. 
The most probable topology ($T1$) has a high support (0.7).
The archaea tree ($T2$) has a posterior probability of 0.172 while the support for the eocyte tree is less than 0.001 (not shown). 
Unlike BayesCAT, the traditional sequential approach strongly supports the split for archaeal monophyly (posterior probability $>$ 0.999).

\begin{table}[t]
 \caption[{\bf 5S rRNA : Summary of posterior distributions of the topology.}]{\label{5srRNAtree:table} {\bf 5S rRNA : Summary of posterior distributions of the topology. } 
 $T1$, $T2$, and $T3$ are the top three topologies ranked by posterior probabilities from BayesCAT.  The full name corresponding to each abbreviation can be found in Table~\ref{5srRNAdata:table}. Archaea taxa are shown in boldface.} 
\begin{center}
\vspace*{0.5cm}   
\includegraphics[scale=0.7]{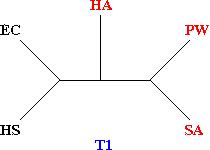}
\hspace*{0.2cm}
\includegraphics[scale=0.83]{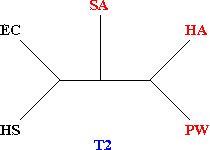}
\hspace*{0.2cm}
\includegraphics[scale=0.83]{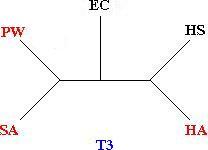}
\\\vspace*{0.5cm}   
    \begin{tabular}{|c|c|c|c|c|}
      \hline
      Method&$T1$&$T2$&$T3$&EC,HS $|$ {\bf HA,PW,SA}\\ \hline \hline
      BayesCAT&0.205&0.170&0.130&0.414\\\hline
      BAli-Phy&0.284&0.103&0.189&0.418\\\hline
      MrBayes+ClustalW&0.700&0.172&$<$0.001&$>$0.999\\\hline
    \end{tabular}
  \end{center}
\end{table}   

To compare BayesCAT to another joint model, posterior probabilities from BAli-Phy are also presented in Table~\ref{5srRNAtree:table}. 
We note that these posterior probabilities are not identical to results in \citet{Redelings05} as the version of BAli-Phy used here improves upon the version used in their publication.
Although BAli-Phy assumes a different indel model, its results are very similar to that from BayesCAT. 
The top three topologies ranked by posterior probabilities from BAli-Phy also are $T1$, $T2$ and $T3$, although $T3$ has higher posterior probability than $T2$ with the BAli-Phy model.  
Like BayesCAT, no single topology is strongly supported by BAli-Phy. 
Of particular note, the posterior probabilities of the hypothesis of archaeal monophyly in BAli-Phy (0.418) and BayesCAT (0.414) are very similar.

In summary, the traditional sequential analysis based on the 5S rRNA sequences leads to strong posterior support for a single topology consistent with the hypothesis of archaeal monophyly, but the joint model moderates this strong support by considering alignment uncertainty. In addition, we observe that two different joint models yield quite similar support for the hypothesis of archaeal monophyly.

\subsection{Summary of alignment samples}
\label{Results:5srRNAalign}
\begin{figure}[t]
\centering
\includegraphics[scale=0.7]{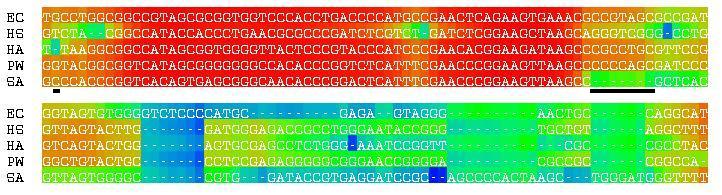}
\\\vspace*{0.2cm}
{\small(a) BayesCAT}
\\\vspace*{0.2cm}
\includegraphics[scale=0.7]{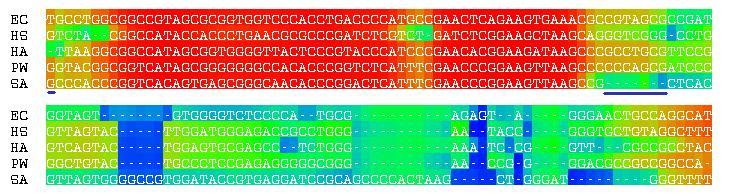}
\\\vspace*{0.2cm}
{\small(b) BAli-Phy}
\\\vspace*{0.2cm}
\includegraphics[scale=1.0]{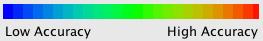}
\\\vspace*{0.5cm}
\caption[{\bf 5S rRNA : Summary of alignment samples.}]{\label{5srRNAalign:fig}
{\bf 5S rRNA : Summary of alignment samples. } Alignment samples from BayesCAT (a) and BAli-Phy (b) are summarized using the procedure described in Section~\ref{method:AlignSummary}}
\vspace*{0.3cm}
\hrule
\end{figure}

Alignment samples from BayesCAT and BAli-Phy are summarized in Figure~\ref{5srRNAalign:fig} using the procedure described in Section~\ref{method:AlignSummary}.
Although BAli-Phy provides its own summarization method, we use the same summarization procedure for both programs to focus the comparison on the alignment distributions and not the summarization methods.
The two point estimates under the different joint models have the same columns in the first half of the alignment except for positioning of two gaps (underlined in Figure~\ref{5srRNAalign:fig}). 
In addition, red color in the first half of the columns indicates that the two point estimates have high expected accuracy under each model.     
In contrast, in the second half of the alignment, the two point estimates are quite different and also show low expected accuracy, illustrated by the blue color.
Most of the gaps observed in both alignments are not shared by multiple taxa, which is consistent with the explanation that most of the indel events happen on the external edges (Section~\ref{Results:5srRNAothers}).

To investigate differences in the alignment distribution between BayesCAT and BAli-Phy, we plot the pairwise homology posterior probabilities from each method (see Web Figure 3 (a)). 
Points form a broad band around a diagonal, but no point is substantially far from the diagonal.
To compare with variability from Monte Carlo error, we plot the pairwise homology probabilities from two different MCMC samples using BayesCAT (see Web Figure 3 (b)).
As the two plots show a similar deviation from the diagonal, the difference in the alignment distribution samples between the two methods may be due in large part to Monte Carlo error. 

The multiple sequence alignment estimated using FSA is given in Web Figure 4. The beginning of the alignment, colored in red, is very similar to that of point estimates under joint models in terms of an alignment of residues as well as their expected accuracy, but very little similarity is observed in the remaining part. 

\subsection{Information on the indel process}
\label{Results:5srRNAothers}
Since we model indel events directly, some information about the indel process can be inferred by using our approach and not by BAli-Phy.
Web Figure 5 shows the posterior estimate of realized indel fragment size distribution, which is obtained by first collecting empirical indel fragment size distributions from each sample, and then averaging over all samples.
This distribution has modes at sizes one and seven.

\begin{table}[t]
\caption[{\bf 5S rRNA : Posterior mean number of indel events on each split (BayesCAT).}]{\label{5srRNANumindel:table}{\bf 5S rRNA : Posterior mean number of indel events on each split (BayesCAT). } The second column lists posterior probabilities for each split. The posterior means of the number of indel events and the edge length given occurrence of each split are shown in the third and fourth columns, respectively. Archaea taxa are shown in boldface.}
  \begin{center}
    \begin{tabular}{|c|c|c|c|}
      \hline
      Split & PP of split& \# of indels&edge length   \\ \hline \hline
      EC $|$ HS,{\bf HA,PW,SA} & 1&2.3&0.456            \\ \hline
      HS $|$ EC,{\bf HA,PW,SA} & 1&3.5&0.464            \\ \hline 
      {\bf HA} $|$ EC,HS,{\bf PW,SA} & 1&2.7&0.264      \\ \hline 
      {\bf PW} $|$ EC,HS,{\bf HA,SA} & 1&0.8&0.147      \\ \hline 
      {\bf SA} $|$ EC,HS,{\bf HA,PW} & 1&3.3&0.366      \\ \hline 
      EC,HS $|$ {\bf HA,PW,SA} & 0.41&0.16&0.112        \\ \hline 
      EC,{\bf HA} $|$ HS,{\bf PW,SA} & 0.08&0.33&0.046  \\ \hline 
      EC,{\bf PW} $|$ HS,{\bf HA,SA} & 0.12&0.11&0.080  \\ \hline 
      EC,{\bf SA} $|$ HS,{\bf HA,PW} & 0.16&0.55&0.076  \\ \hline 
      HS,{\bf HA} $|$ EC,{\bf PW,SA} & 0.35&0.38&0.112  \\ \hline 
      HS,{\bf PW} $|$ EC,{\bf HA,SA} & 0.007&0&0.025    \\ \hline 
      HS,{\bf SA} $|$ EC,{\bf HA,PW} & 0.161&0.37&0.097 \\ \hline 
      {\bf HA,PW} $|$ EC,HS,{\bf SA} & 0.303&0.04&0.090 \\ \hline 
      {\bf HA,SA} $|$ EC,HS,{\bf PW} & 0.053&0.26&0.044 \\ \hline 
      {\bf PW,SA} $|$ EC,HS,{\bf HA} & 0.38&0.41&0.105  \\ \hline 
    \end{tabular}
  \end{center}
\end{table}   

Another quantity we can estimate is the number of indel events on each split.
Table~\ref{5srRNANumindel:table} shows the posterior mean of number of indel events given occurrence of each split.
Most splits corresponding to external edges (the first five splits in Table~\ref{5srRNANumindel:table}) include more than one indel event while the mean number of indel events on internal edges are fewer than one. 
Occurrence of indel events on internal edges implies that two or three leaves (in the five-taxon case) share these events.
Thus, the fact that most of the indel events are observed on the external edges supports that the 5S rRNA sequences we examined do not have a strong phylogenetic signal for shared indel events.

To investigate whether the expected number of indel events vary with branch length, we also list the posterior mean of edge length given occurrence of each split in the fourth column of Table~\ref{5srRNANumindel:table}. 
Edges with more than one indel event are longer than the remaining edges. 
The second split has the largest number of indel events (3.5) on the longest edge length (0.464) while the split with no indel events has the shortest edge length (0.025).

\subsection{Convergence}
\label{convergence:5SrRNA}
We run three MCMC chains from different starting points. Each run has 1,000,000 iterations and we sampled every 1000 iterations. 
To assess convergence for continuous parameters, we compute Gelman-Rubin R statistics \citep{Gelman1992} for sampled external branch lengths and substitution and indel parameters. 
All statistics are less than 1.05 which is consistent with convergence.
Convergence for the tree topology is evaluated as follows. 
For each clade which appeared in any of the three runs, we calculate the relative frequency with which the clade occurs in each of the runs. 
Differences between the minimum and maximum values of these relative frequencies over three runs are less than 5$\%$ for all clades.

\section{Discussion}
\label{ch:discuss}
We have developed a joint model for co-estimation of the alignment and tree. 
Our general indel model allows arbitrary-length overlapping indels and a general distribution for the indel fragment size.
We designed and implemented MCMC methods to carry out Bayesian inference of multiple sequence alignment and phylogeny on the basis of this model. 
Our method for joint estimation improves estimates from the traditional sequential approach by accounting for uncertainty in the alignment in phylogeny inferences, which is demonstrated by real data and simulated data (results from simulated data is given in Web Appendix J). 

Our method is the first approach which includes a complete history of indel events mapped onto the tree as the state space in the Markov Chain.
A large state space containing the complete history of indel events makes our MCMC approach more challenging, but it enables us to infer more information about the indel process itself than can be done with alternative joint model approaches.
Inferred information about the indel process has the potential to be very valuable for some questions of biological interest for some data sets.
In addition to quantities presented in this paper, we can infer more information, e.g., positions of indel events and the proportion of overlapping indels.
Our method would be useful to a biologist interested in the indel process itself.

The alternative approaches sum over all possible indel histories, which places severe constraints on the choice of indel models, e.g., distribution for indel fragment size and number of indel events. 
Thus, our method has the advantage of being relatively easy to extend to model more closely real processes of insertion and deletion. 

To summarize alignment samples, we suggest using a method implemented in FSA and describe how to use it in the joint estimation setting. 
Although this suggestion improves on alternative summarization methods in terms of estimating a point-estimate and showing uncertainty in the point-estimate, the information provided by this summary is still limited. 
We also investigate pairwise posterior probabilities of homology, but our investigation is still ad hoc and cannot provide enough information to fully summarize alignment uncertainty.
Thus, developing methods to present alignment distribution in a more informative manner remains an interesting open challenge. 

Although our method has several advantages relative to the alternative approaches, it still has points which need to be investigated more thoroughly. 
We have not observed notable differences in inferences between our method and BAli-Phy in data analysis although our method assumes a more general indel model. 
We need to investigate when our method can have some advantages. 
Such advantages would presumably be most likely to occur if the true history is likely to contain overlapping indel events. 

\section*{Acknowledgements}
We thank C\'ecile An\'e, Colin Dewey, David Baum, and Michael Newton for helpful suggestions.
We thank the authors of FSA for modifying their software for our use, and the authors of BAli-Phy for sharing their research experience in joint estimation of alignment and tree.


\section*{Supplementary Materials}

Web Appendix, Web Figure, and Web Table are available with
this paper \\at https://github.com/heejungshim/BayesCAT/tree/master/doc/paper. \vspace*{-8pt}

\bibliographystyle{rss} 
\bibliography{BayesCAT}

\end{document}